\documentclass[english,aps,superscriptaddress,prl,notitlepage,reprint]{revtex4-1}
\usepackage{natbib}
\usepackage{babel}
\usepackage{amsmath}
\usepackage{amssymb}
\usepackage{bm}
\usepackage{braket}
\usepackage{graphicx}
\usepackage{comment}
\usepackage{hyperref}
\usepackage{color}
\usepackage{epsfig,natbib}
\newcommand{\be}{\begin{equation}}
\newcommand{\ee}{\end{equation}}
\newcommand{\bea}{\begin{eqnarray}}
\newcommand{\eea}{\end{eqnarray}}
\newcommand{\aop}[1]{{a_{#1}}}
\newcommand{\aopd}[1]{{a^{\dagger}_{#1}}}
\hypersetup{
    colorlinks=true,
    linkcolor=blue,
    citecolor=blue,    
    urlcolor=blue,
}
\begin{document}
\title{Asymptotic Freedom and Large Spin Antiferromagnetic Chains}
\author{Samuel Gozel}
\author{Fr\'ed\'eric Mila}
\affiliation{Institute of Physics, \'Ecole Polytechnique F\'ed\'erale de Lausanne (EPFL), CH-1015 Lausanne, Switzerland}
\author{Ian Affleck}
\affiliation{Department of Physics and Astronomy and Stewart Blusson Quantum Matter Institute, University of British Columbia, 
Vancouver, British Columbia, Canada, V6T1Z1}
\begin{abstract}
Building on the mapping of large-$S$ spin chains onto the O($3$) nonlinear $\sigma$ model with coupling constant $2/S$, and on general properties of that model (asymptotic freedom, implying that perturbation theory is valid at high energy, and Elitzur's conjecture that rotationally invariant quantities are infrared finite in perturbation theory), we use the Holstein-Primakoff representation to derive analytic expressions for the equal-time and dynamical spin-spin correlations valid at distances smaller than $S^{-1} \exp(\pi S)$ or at energies larger than $J S^2 \exp(-\pi S)$, where $J$ is the Heisenberg exchange coupling. This is supported by comparing the static correlations with quantum Monte Carlo simulations for $S = 5/2$.
\end{abstract}
\maketitle

Spin chains have been the subject of intensive investigation since the early days of quantum mechanics. Thanks to the Bethe ansatz, bosonization, and numerical simulations, the static and dynamical properties of the spin-$1/2$ chain are very well understood~\cite{bethe_theorie_1931,des_cloizeaux_spin-wave_1962,karbach_two_1997,caux_four_2006,giamarchi_quantum_2003}. There is no exact solution for the spin-$1$ chain, but thanks to the mapping onto the O($3$) nonlinear $\sigma$ model initiated by Haldane and to extensive quantum Monte Carlo (QMC) simulations and time-dependent density matrix renormalization group (DMRG) results, spin-$1$ chains are fairly well understood as well~\cite{haldane_nonlinear_1983,haldane_continuum_1983,white_spectral_2008}. In comparison, much less is known for spin chains with spin $S\geq 3/2$. According to the mapping onto the nonlinear $\sigma$ model, half-integer spin chains are expected to be gapless while integer spin chains are expected to be gapped. The difference can be traced back to the presence of a topological term in the case of half-integer spins. The critical theory of the $S=1/2$ chain is known to be the ${\rm SU}(2)_1$ Wess-Zumino-Witten conformal field theory and all higher half-odd-integer spins are expected to lie in the same universality class~\cite{affleck_critical_1987,affleck_critical_1989,hallberg_critical_1996}. Alternatively the presence of a gap for integer spin chains up to $S=4$ has been confirmed by QMC simulations~\cite{todo_cluster_2001,todo_parallel_2019}, but there is no definitive information on the dynamics of large-$S$ spin chains because of the difficulty to extend time-dependent DMRG to large spin. 

Another very useful concept for analyzing spin chains is asymptotic freedom, a concept borrowed from quantum chromodynamics according to which quarks appear as nearly free particles in certain high-energy experiments because the effective coupling constant is small at high energy~\cite{gross_ultraviolet_1973,politzer_reliable_1973,peskin_introduction_1995}. For integer spin chains, which are simply described by the O($3$) nonlinear $\sigma$ model, the gap corresponds to the energy scale at which the coupling constant becomes of order 1 as we reduce the energy (see below). Interestingly enough, the more fundamental aspect of asymptotic freedom, namely the fact that perturbation theory can be used at high energy (or at short distance), has not been used to discuss the properties of spin chains. 

In this Letter, we show how to derive analytical results for large-$S$ spin chains using perturbation theory in 1/S for the equal-time correlation function and the dynamical structure factor defined by
\bea
G(|j|) & \equiv & \braket{\bold{S}_j\cdot \bold{S}_0},\nonumber \\
S(k,\omega) & \equiv & \sum_j e^{-i k j} \int_{-\infty}^\infty {\rm d}t \,  e^{i\omega t} \braket{\bold{S}_j(t)\cdot \bold{S}_0(0)}.
\eea
This relies on two fundamental observations: (i) the domain of validity of these results, set by asymptotic freedom, extends to an energy that is exponentially small with $S$ (or to distances that are exponentially large with $S$), so that already for $S=5/2$ most of the parameter space in energy and distance is covered by this perturbative approach. (ii) Divergences typical of quantum fluctuations in 1D often cancel out when averaging over direction.

Let us first review the connection between spin chains and the nonlinear $\sigma$ model, and the consequences of asymptotic freedom. The low-energy (compared to $J S$) degrees of freedom of the antiferromagnetic Heisenberg chain of spin $S$, with Hamiltonian
\be
H = J\sum_j \bold{S}_j \cdot \bold{S}_{j+1},
\ee
can be mapped into those of the O($3$) nonlinear $\sigma$ model (NL$\sigma$M) with Lagrangian density~\cite{haldane_nonlinear_1983,haldane_continuum_1983}
\be
{\cal L} = \frac{1}{2 g} \left( \frac{1}{v} (\partial_t \bm{\phi} )^2 - v (\partial_x\bm{\phi} )^2\right) + \frac{\theta}{8\pi} \, 
\epsilon^{\mu \nu} \bm{\phi} \cdot (\partial_\mu \bm{\phi} \times \partial_\nu \bm{\phi} ).
\ee
Here the field $\bm{\phi}$ is a three-component unit vector, $|\bm{\phi} (t,x)|^2=1$.  The last term is topological;
it has no effect in perturbation theory. 
The parameters take the approximate values (setting the lattice constant $a$ of the spin chain to $1$),
\be
g = \frac{2}{S}, \quad v = 2 J S, \quad \theta =2\pi S.
\label{equ::def_g_def_v_def_theta}
\ee
This approximate mapping is obtained using
\be
\bold{S}_j \approx (-1)^j S \, \bm{\phi}(x_j) + \bm{l}(x_j),
\label{mapping1}
\ee
where 
\be
\bm{l} = \frac{1}{v g} \, \bm{\phi} \times \partial_t\bm{\phi}
\label{mapping2}
\ee
is the conserved spin density of the $\sigma$ model. The field $\bm{\phi}$ is the antiferromagnetic order parameter.

A perturbative treatment of the $\sigma$ model assumes spontaneous breaking of the O($3$) symmetry with, for 
example, $\bm{\phi} \approx (0,0,1)$. We then introduce a pair of Goldstone boson fields, $\bm{\varphi} = (\varphi^1$, $\varphi^2)$ and write
\be
\bm{\phi} =(\varphi^1,\varphi^2,\sqrt{1-\bm{\varphi}^2}).
\ee
Rescaling $\bm{\varphi} \to \sqrt{g} \, \bm{\varphi}$, and setting the velocity to 1, 
the nontopological part of the Lagrangian density becomes
\be
{\cal L} = \frac{1}{2} \left[  \partial^\mu \bm{\varphi} \cdot \partial_\mu \bm{\varphi} +
\frac{g (\bm{\varphi} \cdot \partial^\mu \bm{\varphi})(\bm{\varphi} \cdot \partial_\mu \bm{\varphi} )}{1 - g \bm{\varphi}^2} \right].
\ee
Taylor expanding the denominator in the second term gives a series of interaction terms. But perturbation theory is plagued with infrared divergences in (1+1) dimensions. This is related to the Mermin-Wagner-Coleman theorem~\cite{mermin_absence_1966,coleman1973}. This failure of perturbation theory can be understood using the renormalization group, just as in QCD. The renormalized coupling constant at length scale $L$ is determined by~\cite{polyakov_interaction_1975}
\be
\frac{{\rm d}g}{{\rm d}\ln L} = \frac{1}{2\pi} g^2 + \frac{1}{(2\pi)^2} g^3 + \mathcal{O}(g^4).
\ee
Integrating this equation and using the expression of the bare coupling constant $g$ given in Eq.~\eqref{equ::def_g_def_v_def_theta} we obtain the crossover length scale at which $g(L)$ becomes $\mathcal{O}(1)$ (see, for instance Ref.~\onlinecite{fradkin_field_2013} and references therein),
\be
\xi \propto \frac{1}{S} e^{\pi S} ( 1 + \mathcal{O}(S^{-1}))
\ee
This implies that the effective coupling constant becomes $\mathcal{O}(1)$ at the energy scale $\Lambda = v/\xi \propto J S^2 e^{-\pi S} (1 + \mathcal{O}(S^{-1}))$.
For integer spin, corresponding to $\theta =0$, there is a ``Haldane gap" of order $\Lambda$.  For half-integer spin,
corresponding to $\theta =\pi$, the model is gapless but $\Lambda$ is nonetheless a crossover scale. 
At energy scales $\gg \Lambda$ or, correspondingly length scales $\ll \xi \equiv v/\Lambda$, we might expect perturbative 
behavior to hold, corresponding to asymptotic freedom. This behavior will be the same for both integer and half-integer spin. The field theory approximation only holds at energy scales small compared to the bandwidth, $J S$.  But, for large $S$, 
$\Lambda \ll J S$ so there is an intermediate energy window, $\Lambda \ll E \ll J S$, in which a perturbative treatment of the  $\sigma$ model applies. 
As we go to larger $S$ the lower bound of this energy window extends almost to zero. Table~\ref{table::energyscales} summarizes the values of $\Lambda$ and $\xi$ obtained from numerical measurement of the gap for integer spin $S = 1,2,3$, as well as the expected values for $S = 3/2$ and $5/2$ obtained by interpolation~\cite{todo_cluster_2001}.

\begin{table}
\begin{ruledtabular}
\begin{tabular}{cccccc}
$S$ & $1$ & $3/2$ & $2$ & $5/2$ & $3$ \\
\colrule
$\Lambda/J$ & $0.410$ & $\sim  0.1$ & $0.0892$ & $\sim 0.02$ & $0.0100$ \\
%--------------
$\xi$ & $6.02$ & $\sim 14$ & $49.5$ & $\sim 160$ & $637$ \\
\end{tabular}
\end{ruledtabular}
\caption{Values of the energy scale $\Lambda$ and associated length scale $\xi$ for the relevant values of the spin $S$.}
\label{table::energyscales}
\end{table}

Perturbation theory works at high energies in the $\sigma$ model for rotationally invariant quantities only. 
In particular, such angular averaging leads, as we will see, to cancellations of infrared divergences, resulting 
in infrared-finite perturbative results. This is known as ``Elitzur's conjecture"~\cite{elitzur_applicability_1983,david_cancellations_1981}. 
Indeed, the staggered part of the equal-time spin chain Green's function is given by the $\phi^a$ Green's function.  Up to $\mathcal{O}(g)$ this gives
\bea
\braket{\phi^a(x) \phi^b(0)} &\approx & g \braket{\varphi^a(x)\varphi^b(0)}, \quad (a,b\in \{1,2\})\nonumber \\
%------------
\braket{\phi^3(x) \phi^3(0)} & \approx & 1 - g \braket{\bm{\varphi}(0)\cdot \bm{\varphi}(0)}.
\eea
The free massless boson Green's function is
\be
\braket{\varphi^a(x)\varphi^b(0)} = \delta^{ab}\int {{\rm d}k\over 4\pi |k|}e^{i k x}.
\ee
We may insert an ultraviolet cutoff, $|k|<D/v$, where $D\approx J S$.  However, there is also a logarithmic infrared divergence at $k\approx 0$. 
Thus, $\braket{\phi^a(x)\phi^b(0)}$ is infrared divergent. Averaging over directions leads to
\be
\braket{\bm{\phi}(x)\cdot \bm{\phi}(0)} \approx 1+g\int {{\rm d}k\over 4\pi|k|} \left (e^{i k x} - 1 \right),
\ee
which is an infrared-finite result at all energy scales.

Performing the integration leads to the well-known result~\cite{elitzur_applicability_1983}
\be 
\braket{\bm{\phi}(x)\cdot \bm{\phi}(0)} \approx 1-{g \over \pi}\ln (|x|D/v).
\ee
In order to use the $\sigma$ model to make predictions for the spin chain, it is also useful to calculate 
$\braket{l^a(x)l^b(0)}$. Using
\be
\bm{l} \approx \frac{1}{\sqrt{g} \, v} (-\partial_t\varphi^2,\partial_t\varphi^1,0),
\ee
we get
\be
\braket{\bm{l}(x)\cdot \bm{l}(0)} \approx -{1\over \pi gx^2}.
\ee
(In this case, averaging is not necessary to get an infrared-finite result.)

Using the mapping of Eqs.~\eqref{mapping1} and~\eqref{mapping2}, the spin-spin correlation is given by
\be
\braket{S^a_j(t)S^b_0(0)} \approx S^2(-1)^j\braket{\phi^a (t,j)\phi^b (0,0)}+\braket{l^a(t,j)l^b(0,0)}.
\label{equ::NLSM::SajSb0}
\ee
From Eq.~\eqref{equ::NLSM::SajSb0} we see that, in perturbation theory, these Green's functions are very different for $a=b=3$ 
and $a=b=1$ or $2$. In fact, we only expect perturbation theory to be valid if we average over directions, 
using
\be
\braket{S^a_j(t)S^b_0(0)} = {\delta^{ab}\over 3}\braket{\bold{S}_j(t)\cdot \bold{S}_0(0)}.
\ee
Thus, in the $\sigma$ model approximation
we predict
\be
\braket{\bold{S}_j\cdot \bold{S}_0} \approx  (-1)^j S^2 \left[ 1 - \frac{2}{\pi S} \ln\left( \frac{|j| D}{v}\right) \right] - \frac{S}{2\pi j^2}
\label{equ::sigmaModel::spinspincorrelation}
\ee
for $1\ll |j|\ll \xi$ and $S$ integer or half-integer. This expression explicitly breaks down at distance $v \, e^{\pi S/2}/D \approx \xi$. The correct exponent of the exponential factor for this crossover length scale can be obtained from Eq.~\eqref{equ::sigmaModel::spinspincorrelation} by adding second order corrections. For $|j|\gg \xi$ we expect exponential decay 
for integer $S$. For half-integer $S$, $(-1)^j/|j|$ decay is expected.

Now we apply nonlinear spin-wave theory (NLSWT) to the spin chain using the Holstein-Primakoff (HP) representation of spins. The advantage of this perturbative approach on the lattice is that the energy window of the perturbative regime extends to arbitrary high energies, $E \gg \Lambda$. For this reason we expect HP perturbation theory to provide a more accurate description of the spin chain at very short distances.
Using HP transformation we rewrite all spin operators in terms of HP bosons and treat the model perturbatively in $1/S$~\cite{holstein_field_1940}. We begin by computing, to first order, the spin-spin correlation along the $z$ axis:
\be
\braket{S^z_0 S^z_r} = (-1)^r S^2 \left[ 1 + \frac{1}{S} \left( 1 - \frac{2}{\pi} \int_0^{\pi/2} {\rm d} k \, \frac{1}{\sin k} \right) \right].
\label{equ::SzSz_swt}
\ee
The first order term is obviously infrared divergent. The correlations along $x$ and $y$ axis are equivalent and must be computed separately for even and odd distance. We obtain to first order,
\bea
\braket{S^x_0 S^x_r} & = & \frac{S}{\pi} \int_0^{\pi/2} {\rm d} k \, \frac{\cos(k r)}{\sin k}, \quad r \ \text{even},  \label{equ::SxSx_swt_even}\\
%---------
\braket{S^x_0 S^x_r} & = & - \frac{S}{\pi} \int_0^{\pi/2} {\rm d} k \, \frac{\cos(k r)}{\tan k}, \quad r \ \text{odd}. 
\label{equ::SxSx_swt_odd}
\eea
The two integrals involved in these expressions are divergent at $k=0$. However adding Eq.~\eqref{equ::SzSz_swt} to twice Eq.~\eqref{equ::SxSx_swt_even} or~\eqref{equ::SxSx_swt_odd} for even or odd distance, respectively, leads to infrared-finite results. Extending the calculation to second order in perturbation theory to the rotationally invariant spin-spin correlation, we obtain~\cite{SM}
\be
\begin{aligned}
\braket{\bold{S}_0 \cdot \bold{S}_{r}} = & \ (-1)^{r} S^2 \left[ 1 + \frac{1}{S} \left( 1 - \frac{2}{\pi} J_{\alpha}(r) \right) \right. \\
+ & \left. \frac{1}{4 S^2} \left( 1 - \frac{2}{\pi} J_{\alpha}(r) - \delta_{r,0} \right)^2 + \mathcal{O}(S^{-3}) \right]
\label{equ::SWT::spin-spin-correlation}
\end{aligned}
\ee
where $\alpha = r \, (\text{mod} \, 2)$ and where $J_{0,1}(r)$ are the two following infrared-finite integrals:
\bea
J_0(r) &=& \int_0^{\pi/2} {\rm d}k \, \frac{1 - \cos(k r)}{\sin k}, \nonumber \\
J_1(r) &=& \int_0^{\pi/2} {\rm d}k \, \left[ \frac{1}{\sin k} - \frac{\cos(k r)}{\tan k} \right].
\label{equ::integrals_J0J1}
\eea

\begin{figure}
\begin{center}
\includegraphics[width=\columnwidth]{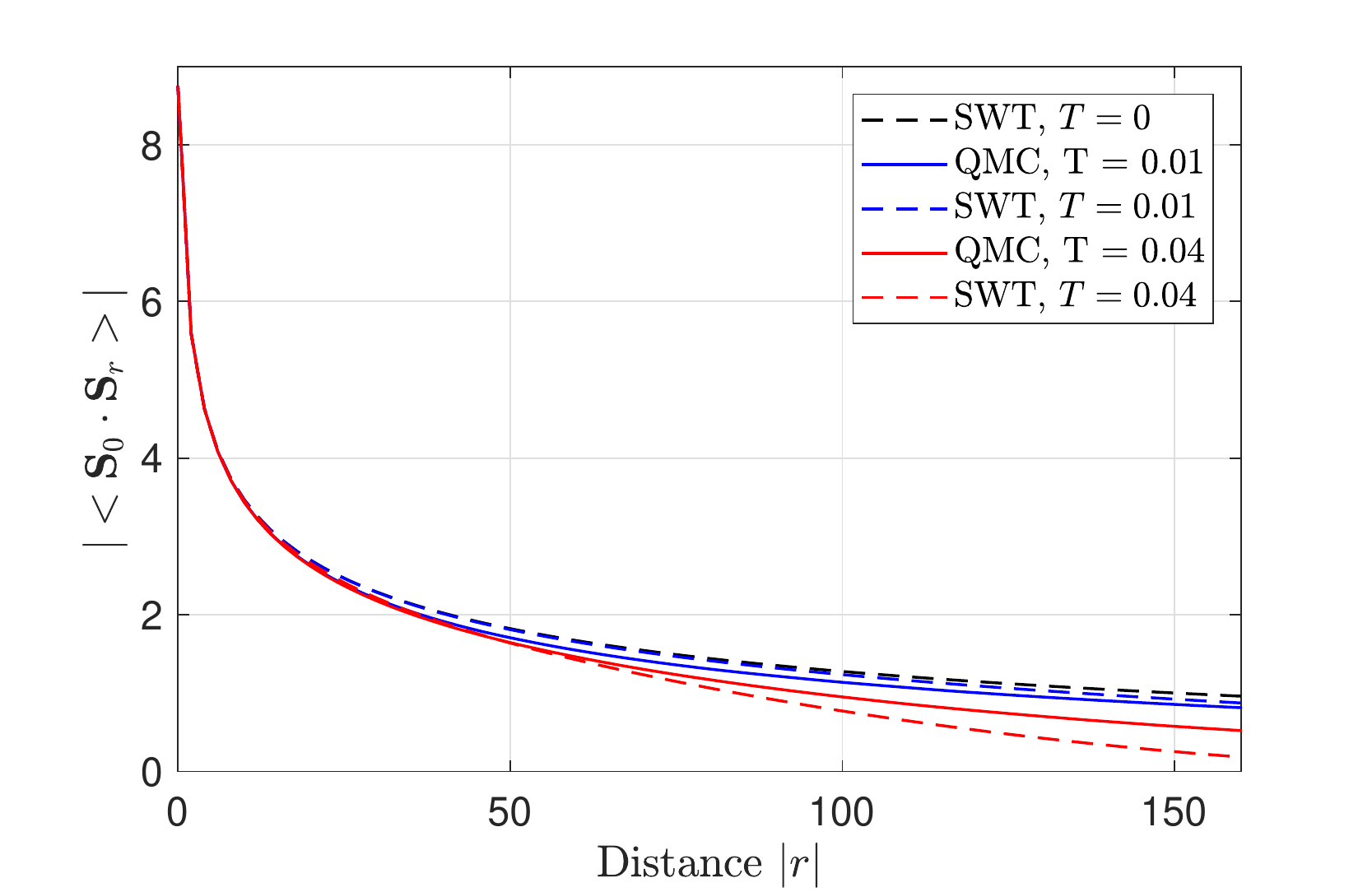}
\end{center}
\caption{Static spin-spin correlation function obtained in NLSWT for $S = 5/2$ at zero and finite-temperature and comparison to QMC~\cite{todo_cluster_2001}. The errorbars are smaller than visible.}
\label{fig::statics::T}
\end{figure}

This formula is thus in agreement with Elitzur's conjecture in the O($3$) NL$\sigma$M. The O($3$) invariant two-point function of spin operators in spin-wave theory is infrared finite to second order. The divergences occurring in $\braket{S^z_0 S^z_{r}}$ are exactly canceled by the divergences of $2\braket{S^{x}_0 S^x_{r}}$~\cite{SM}. Figure~\ref{fig::statics::T} compares Eq.~\eqref{equ::SWT::spin-spin-correlation}, as well as its finite-temperature generalization, to QMC simulations for spin $S = 5/2$ in the regime $|r| < \xi \approx 160$. At finite temperature the agreement between the perturbative calculation and the QMC result is excellent up to distance $|r| \simeq 50 \simeq \xi/3$ and goes beyond the initial expectations of validity $|r| \ll \xi$. Taking now the limit of large distance $|r| \gg 1$ but always keeping $|r| \ll \xi$ we obtain
\be
\begin{aligned}
\braket{\bold{S}_0 \cdot \bold{S}_{r}} \simeq & \ (-1)^{r} S^2 \left[ 1 - \frac{2}{\pi S} \ln\left(\frac{|r|}{r_0}\right) + \frac{1}{2\pi S} \frac{1}{r^2} \right. \\
& \hspace{1.6cm} \left. + \frac{1}{\pi^2 S^2} \ln^2\left(\frac{|r|}{r_0}\right) \right] - \frac{S}{2\pi r^2}
\end{aligned}
\label{equ::SdotS_longdist}
\ee
where $r_0 = e^{\pi/2-\gamma}/2 \simeq 1.35$ and $\gamma$ is the Euler-Mascheroni constant~\cite{SM}. This is in agreement with the NL$\sigma$M. Spin-wave theory thus predicts a logarithmic decay of the spin-spin correlation function at short distance with respect to the crossover length scale $\xi$, very different from the exponential or power-law decay at long distance $|r| \gg \xi$. The $1/r^2$ term at order $1/S$ in the staggered part of Eq.~\eqref{equ::SdotS_longdist} comes from small violation of Lorentz invariance in spin-wave theory and is subdominant at large distance $1 \ll |r| \ll \xi$ compared to the logarithmic decay.

The dynamical structure factor can be obtained from the time-ordered Green's function of HP bosons,
\be
i \, G(k,t) = \braket{T \begin{bmatrix}
\aop{k}(t) \\ \aopd{-k}(t)
\end{bmatrix}
\begin{bmatrix}
\aopd{k}(0) & \aop{-k}(0)
\end{bmatrix}}
\label{equ::SWT::G_HP}
\ee
and its Fourier transform $G(k,\omega)$ as~\cite{doniach_greens_1974}
\be
S^{ab}(k,\omega) = - 2 \, \text{Im}\left[ F^{ab}(k,\omega) \right]
\ee
where $F^{ab}(k,\omega)$ is the time-ordered Green's function of spin operators,
\be
i \, F^{ab}(k,\omega) = \int_{-\infty}^{\infty} {\rm d}t \, e^{i\omega t} \, \braket{T S^{a}_k(t) S^{b}_{-k}(0)}.
\ee
We get
\be
S^{xx}(k,\omega) = S \pi \left( 1 - \frac{n}{S} \right) \left| \tan\left(\frac{k}{2}\right)\right| \delta(\omega - \epsilon_k)
\label{equ::Sxx_T0}
\ee
where $\epsilon_k$ is the dispersion relation of the spin waves~\cite{SM}.  The coefficient $n=\braket{\aopd{i}\aop{i}}$ is infrared divergent. The transverse part of the dynamical structure factor is thus divergent. First order interaction terms do not broaden the delta peak. The weight is turned from finite and positive at zeroth order to  infinite and negative at first order. This is, however, not a major issue since perturbation theory is only expected to be valid at energies $|\omega^2 - \epsilon_k^2| > \Lambda^2$, in analogy to the NL$\sigma$M where the two-momentum must satisfy $|Q^2|>\Lambda^2$. The longitudinal component of the structure factor is given by a two-magnon continuum starting at energy $\epsilon_k$ and extending up to energies of the order of $4 J S$. The spectral weight diverges quadratically in the frequency close to momentum $k \approx 0,\pi$. Again we only trust the results above the threshold where the spectral weight is finite.

As a consequence of the divergences of $S^{xx}(k,\omega)$ and $S^{zz}(k,\omega)$ the associated components of the static structure factor obtained as
\be
S^{aa}(k) = \int_0^{\infty} \frac{{\rm d}\omega}{2\pi} \, S^{aa}(k,\omega)
\label{equ::Saak}
\ee
are also divergent. However, averaging over directions leads to an infrared-finite result, thus providing a reliable description of the isotropic quantity for $k \gg \Lambda/v$~\cite{SM}. The same result can be obtained by Fourier transforming the static spin-spin correlation function in Eq.~\eqref{equ::SWT::spin-spin-correlation}.

\begin{figure}
\begin{center}
\includegraphics[scale=0.4]{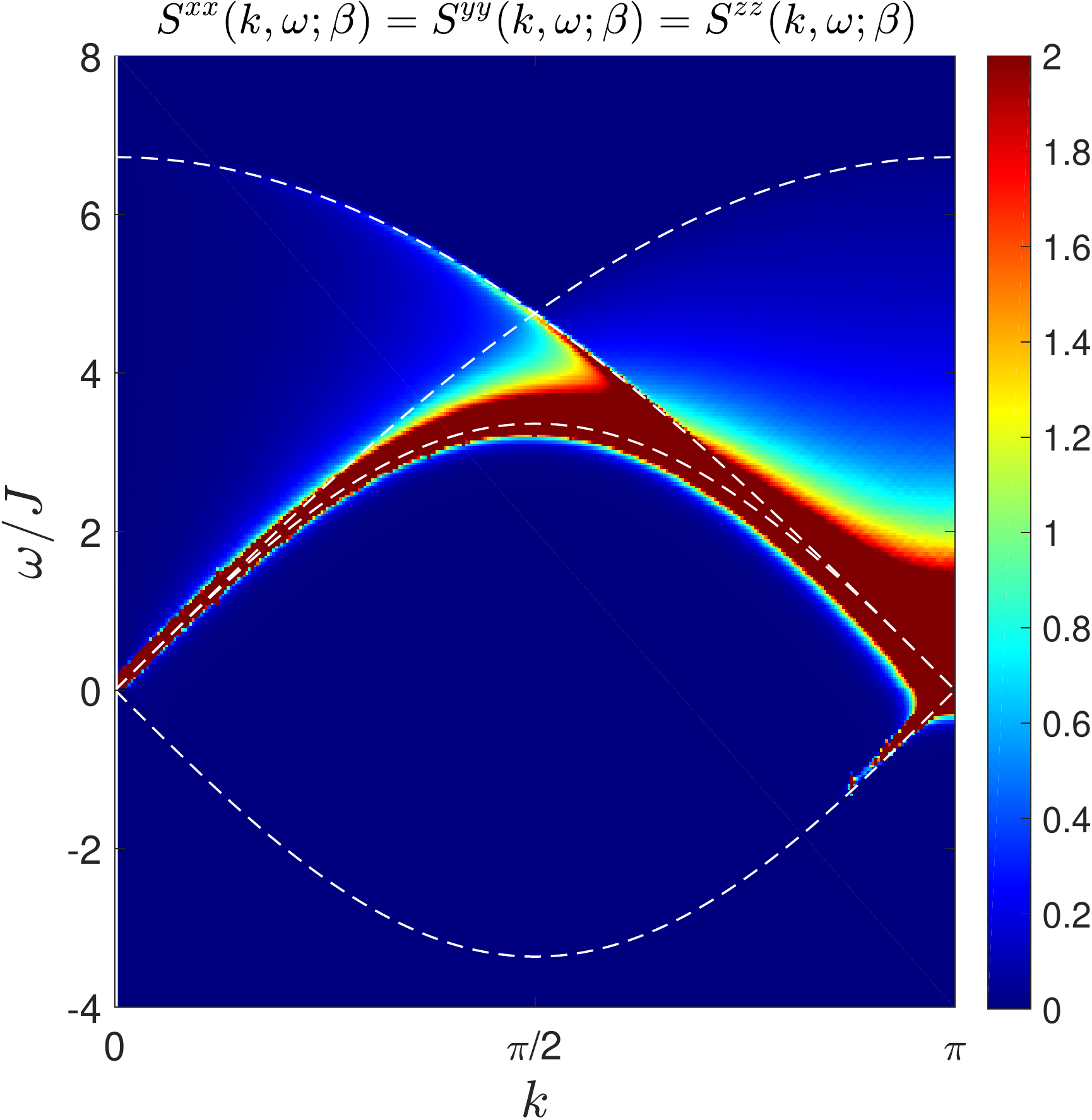}
\end{center}
\caption{Dynamical structure factor for $S = 5/2$ at inverse temperature $\beta = 10$ (in units of $J^{-1}$). The white dashed lines are the lower and upper thresholds of the continuum, $\omega_k = \pm v_{\beta} |\sin(k)|$, $\omega_k = 2 v_{\beta} \sin(k/2)$ and $\omega_k = 2 v_{\beta} |\cos(k/2)|$. Because of temperature effects there is nonzero spectral weight at $|\omega| < v_{\beta} |\sin(k)|$.}
\label{fig::DSF::T}
\end{figure}

One can perform the same calculation at finite temperature $T = 1/\beta$~\cite{SM}. The main tool is now the bosonic Matsubara Green's function $\mathbb{G}(k,i\omega_n)$ defined equivalently to Eq.~\eqref{equ::SWT::G_HP}, and the dynamical structure factor is obtained as~\cite{doniach_greens_1974}
\be
S^{ab}(k,\omega;\beta) = - \frac{2}{1-e^{-\beta\omega}} \,  \text{Im}\left[ F^{ab,R}(k,\omega;\beta) \right]
\ee
where $F^{ab,R}(k,\omega;\beta)$ is the retarded Green's function of spin operators and is obtained from the imaginary-time ordered Green's function by analytical continuation. Since $F^{ab,R}(k,\omega;\beta)$ can be nonzero at negative energy, we expect nonvanishing spectral weight at $\omega<0$. This corresponds to transitions between single-magnon states, as shown in Fig.~\ref{fig::DSF::T}, where we averaged over directions. Along the lower thresholds $|\omega| = v_{\beta} |\sin(k)|$, where $v_{\beta}$ is the temperature-dependent NLSWT velocity, the spectral weight is divergent. However the behavior along the thresholds will in reality be different for integer and half-integer spin: for integer spin, the gap $\Lambda$ appears at momentum $\pi$ and a two-magnon continuum starts at momentum $0$ and energy $2\Lambda$~\cite{white_spectral_2008}. For half-integer spin the spectrum is gapless at momentum $0$ and $\pi$~\cite{des_cloizeaux_spin-wave_1962}. As $S$ increases the spectral weights for integer or half-integer spin well above $\Lambda$, which is exponentially small in $S$, become identical and are described by the perturbative approach developed here.

To summarize, we have shown that reliable perturbative results in $1/S$ can be obtained for large-$S$ spin chains in spite of the infrared divergences typical of quantum field theory or spin-wave expansion in $(1+1)$ dimensions. The static spin-spin correlation has a characteristic logarithmic decay with distance below a crossover length scale that grows extremely fast with $S$ as $S^{-1} e^{\pi S}$, and the dynamical structure factor can be accurately determined except in a very narrow energy window of order $J S^2 e^{-\pi S}$. This perturbative regime cannot be observed in the well-studied cases of spin-$1/2$ and spin-$1$ chains because the crossover length is too small. In that respect, the spin-$5/2$ case illustrated in this Letter is of particular interest because the crossover length scale $\xi\simeq 160$ is large enough to observe an extended perturbative regime and because several compounds with half filled $d$-orbitals realize isotropic quasi-one-dimensional Heisenberg antiferromagnets, allowing for experimental investigation~\cite{dingle_linear-chain_1969,birgeneau_spin_1971,hutchings_spin_1972,mole_non-classical_2009,asha_magnetic_2015}. We hope that the present results will motivate such investigations.\\

\begin{acknowledgments}
S.G. is grateful to Francisco Hyunkyu Kim for helpful discussions and to Synge Todo for help setting up the QMC simulations. The calculations have been performed using the facilities of the Scientific IT and Application Support Center of EPFL. This work has been supported by the Swiss National Science Foundation (SNF), by NSERC Discovery Grant
04033-2016 and by the Canadian Institute for Advanced Research.
\end{acknowledgments}

\bibliography{references}
\bibliographystyle{apsrev4-1}

\end{document}

\title{Supplemental Material \\ Asymptotic Freedom and Large Spin Antiferromagnetic Chains}
\author{Samuel Gozel}
\author{Fr\'ed\'eric Mila}
\affiliation{Institute of Physics, \'Ecole Polytechnique F\'ed\'erale de Lausanne (EPFL), CH-1015 Lausanne, Switzerland}
\author{Ian Affleck}
\affiliation{Department of Physics and Astronomy and Stewart Blusson Quantum Matter Institute, University of British Columbia, 
Vancouver, B.C., Canada, V6T1Z1}

\maketitle

\makeatletter
\renewcommand{\theequation}{S\arabic{equation}}
\renewcommand{\thefigure}{S\arabic{figure}}
\renewcommand{\bibnumfmt}[1]{[S#1]}
\renewcommand{\citenumfont}[1]{S#1}

In this Supplemental Material we provide details on the spin-wave theory calculations. In the first section we introduce our definitions and develop the standard Rayleigh-Schr\"odinger perturbation theory for the Heisenberg chain. In Section~\ref{sec2} we derive Eq.~(23). Section~\ref{sec3} is devoted to the derivation of the asymptotic (large distance) behavior of the integrals given in Eq.~(24). Equation~(25) can then be easily obtained. In Section~\ref{sec4} we extract the static structure factor $S(k)$ at zero temperature from the results obtained in Section~\ref{sec2}. In order to avoid repetitions when treating the finite-temperature case, we postpone the derivation of Eq.~(29) to Section~\ref{sec6} and focus in Section~\ref{sec5} on the calculation of the dynamical spin structure factor at finite temperature $1/\beta$. This contains all the information needed to reproduce Fig.~2. Finally Section~\ref{sec6} deals with the limit $\beta\rightarrow\infty$ to prove Eq.~(29). Moreover, we also apply Eq.~(30) to rederive the expressions for the static spin structure factor at zero temperature obtained in Section~\ref{sec4}.

\section{Holstein-Primakoff perturbation theory\label{sec1}}

We use Holstein and Primakoff representation of spin operators to rewrite the Heisenberg Hamiltonian in terms of bosonic creation and annihilation operators. For convenience we perform a $\pi$-rotation along $x$-axis for all spins living on odd sites $\bold{S}_i \rightarrow \tilde{\bold{S}}_i := (S_i^x,-S_i^y,-S_i^z)^T, \ i \ \text{odd}$ and keep the spin operators on even sites unchanged $\bold{S}_i \rightarrow \tilde{\bold{S}}_i := \bold{S}_i, \ i \ \text{even}$. The transformed spin operators $\tilde{\bold{S}}_i$ on all sites can then be expressed as
\begin{align}
\tilde{S}_i^+ = & \ \sqrt{2 S} \, f_i(S) \, \aop{i}, \label{sm::equ::HP_Sp} \\
\tilde{S}_i^- = & \ \sqrt{2 S} \, \aopd{i} \, f_i(S), \label{sm::equ::HP_Sm} \\
\tilde{S}_i^z = & \ S - a_i^{\dagger} a_i,\label{sm::equ::HP_Sz}
\end{align}
where $\aopd{i}$ ($\aop{i}$) is a creation (annihilation) operator on site $i$ satisfying $[\aop{i}, \aopd{j}] = \delta_{i,j}$ and 
\be
f_i(S) = \sqrt{1 - \frac{\aopd{i}\aop{i}}{2 S}}.
\ee
Assuming $\braket{\aopd{i}\aop{i}} \ll 2 S$ and expanding the square root leads to an infinite sequence of terms in $1/S$ which allows us to rewrite the Hamiltonian as
\be
H = J S \sum_{\braket{ij}} \left[ \aopd{i}\aop{i} + \aopd{j}\aop{j} + \aop{i}\aop{j} + \aopd{i}\aopd{j} \right] + V
\ee
where $V = V_1 + \frac{1}{S} V_2 + ...$ contains all orders in $1/S$ and where we have omitted a constant term. The first order interaction term $V_1$ takes the following form:
\be
V_1 = - \frac{J}{2} \sum_{\braket{ij}} \left( \aopd{i} \aop{i} \aop{i} \aop{j} + \text{h.c.} \right) - J \sum_{\braket{ij}} \aopd{i}\aopd{j}\aop{i}\aop{j}.
\ee

Introducing the Fourier-transformed operators,
\be
\aop{k} = \frac{1}{\sqrt{N}} \sum_j e^{-i k r_j} \aop{j}, \qquad \aopd{k} = \frac{1}{\sqrt{N}} \sum_j e^{i k r_j} \aopd{j}
\ee
with $N$ the number of sites in the chain, $r_j = j a$ and $a$ the lattice spacing, the Hamiltonian becomes
\be
H = H_0 + V
\ee
where
\be
H_0 = \sum_k \left[ A_k \, \aopd{k}\aop{k} - \frac{1}{2} B_k \left( \aopd{k}\aopd{-k} + \aop{k}\aop{-k} \right) \right]
\ee
and
\be
A_k = 2 J S, \qquad B_k = - 2 J S \gamma_k, \qquad \gamma_k = \cos k.
\ee
Similarly the first order interaction $V_1$ becomes
\be
V_1 = \frac{1}{N} \sum_{k_1,k_2,k_3} \Xi_1(k_1) \left( \aopd{k_1+k_2+k_3} \aop{k_1}\aop{k_2}\aop{k_3} + \text{h.c.} \right) + \frac{1}{N} \sum_{k_1,k_2,k_3} \Xi_2(k_1,k_3) \aopd{k_1}\aopd{k_2}\aop{k_3}\aop{k_1+k_2-k_3}
\ee
where the vertices are given by
\be
\Xi_1(k_1) = - \frac{J}{2} \, \gamma_{k_1}, \qquad \Xi_2(k_1,k_3) = - J \gamma_{k_1-k_3}.
\ee
The diagonalization of $H_0$ is obtained with a Bogoliubov transformation where we introduce a new set of bosonic operators through the relations
\be
\alop{k} = u_k \aop{k} + v_k \aopd{-k} \hspace{2cm} \alopd{k} = u_k \aopd{k} + v_k \aop{-k}
\ee
with the standard commutation relations,
\be
\left[ \alop{k}, \alopd{k'} \right] = \delta_{k,k'}.
\ee
Solving one gets
\be
u_k = \sqrt{\frac{A_k + \epsilon_k}{2 \epsilon_k}}, \qquad v_k = - \text{sign}(B_k) \, \sqrt{\frac{A_k-\epsilon_k}{2 \epsilon_k}}
\ee
where the linear spin-wave theory dispersion relation is
\be
\epsilon^{(0)}_k = \sqrt{A_k^2 - B_k^2} = 2 J S |\sin k|.
\ee
Let us now consider the first order interaction $V_1$. It is useful to introduce intermediary quantities:
\begin{align}
\Delta := & \braket{\aop{i}\aop{j}} \hspace{1mm} = - \frac{1}{N} \sum_k \gamma_k u_k v_k \hspace{1.6cm} \text{IR-divergent} \label{equ::def_Delta} \\
%------------------
n := & \ \braket{\aopd{i}\aop{i}} = \frac{1}{N} \sum_k v_k^2  \hspace{2.6cm} \text{IR-divergent} \label{equ::def_n}
\end{align}
where $\braket{\mathcal{O}}$ denotes a vacuum expectation value taken in the free theory: $\braket{\mathcal{O}} = \braket{0|\mathcal{O}|0}$ where $\ket{0}$ is the Bogoliubov vacuum $\alop{k} \ket{0} = 0$. The crucial observation is that
\be
\kappa := n + \Delta = \frac{1}{N} \sum_k v_k \left( v_k - \gamma_k u_k \right) = \frac{1}{N} \sum_k \frac{|\sin k|-1}{2} \longrightarrow \frac{1}{2\pi} \int_{-\pi}^{\pi} \dd k \, \frac{|\sin k|-1}{2} = \frac{1}{\pi} - \frac{1}{2} \simeq -0.181690
\label{equ::def_kappa}
\ee
is infrared finite (we have taken the continuum limit).\\

Using now Wick's decoupling one can show that the first order interaction takes the form:
\be
V_1 = \Gamma_0 + \sum_k \Gamma_1(k) \alopd{k}\alop{k} + :(4-\text{bosons}):
\ee
where $:(4-\text{bosons}):$ denotes normal-ordered terms with $4$ Bogoliubov bosons and where
\be
\Gamma_0 = - J N \kappa^2,
\ee
\be
\Gamma_1(k) = - 2 J \kappa \left[ (u_k^2 + v_k^2) - 2 \gamma_k u_k v_k \right] = - 2 J \kappa |\sin k|.
\ee
$\Gamma_0$ and $\Gamma_1(k)$ renormalize the linear spin-wave theory ground state energy and dispersion relation, respectively, by a finite amount. Another important observation is that $V_1$ does not contain terms with $2$ annihilation or $2$ creation Bogoliubov operators.\\

In what follows we define the NLSWT dispersion relation by
\be
\epsilon_k = \epsilon^{(0)}_k + \Gamma_1(k) = 2 J S \left( 1 + \frac{|\kappa|}{S} \right) |\sin k|.
\label{sm::equ::epsilonk_nlswt}
\ee

\section{Deriving Eq. (23)\label{sec2}}

We turn now to the O($3$) invariant static two-point spin-spin correlation function,
\be
\mathcal{O}_{ij} := \bold{S}_i \cdot \bold{S}_j
\ee
which can be expressed in terms of the HP bosons as a power series in $1/S$. The $\pi$-rotation of odd spins requires to treat separately even and odd distances $|i-j|$ between the two operators. Writing
\be
\mathcal{O}_{ij} = S^2 \left[ \mathcal{O}_{ij}^{(0)} + \frac{1}{S} \mathcal{O}_{ij}^{(1)} + \frac{1}{S^2} \mathcal{O}_{ij}^{(2)} + \mathcal{O}(S^{-3}) \right]
\ee
we obtain
\be
\mathcal{O}_{ij}^{(0)} = (-1)^{i-j},
\ee
\be
\mathcal{O}_{ij}^{(1)} = 
\left\{
\begin{array}{ll}
\aopd{i}\aop{j} + \aop{i} \aopd{j} - \aopd{i} \aop{i} - \aopd{j} \aop{j} & \text{for} \ i-j \ \text{even} \\
%-----------------------
\aopd{i} \aopd{j} + \aop{i} \aop{j} + \aopd{i}\aop{i} + \aopd{j}\aop{j} & \text{for} \ i-j \ \text{odd}, \\
\end{array}
\right.
\label{sm::equ::Oij1}
\ee
\be
\mathcal{O}_{ij}^{(2)} = 
\left\{
\begin{array}{lll}
 & \aopd{i}\aop{i}\aopd{j}\aop{j} - \frac{1}{4} \left[ \aop{i}\aopd{j}\aopd{j}\aop{j} + \aopd{i}\aop{i}\aop{i}\aopd{j} + \aopd{i}\aopd{j}\aop{j}\aop{j} + \aopd{i}\aopd{i}\aop{i}\aop{j} \right] & \text{for} \ i-j \ \text{even} \\
 %-----------------------
- & \aopd{i}\aop{i}\aopd{j}\aop{j} - \frac{1}{4} \left[ \aopd{i}\aop{i}\aop{i}\aop{j} + \aop{i}\aopd{j}\aop{j}\aop{j} + \aopd{i}\aopd{i}\aop{i}\aopd{j} + \aopd{i}\aopd{j}\aopd{j}\aop{j} \right] & \text{for} \ i-j \ \text{odd}. \\
\end{array}
\right.
\ee

Before going through more details we first make the link with Eq.~(20-22) where we have explicitly separated the longitudinal ($S^z_i S^z_j$) and the transverse ($S^x_i S^x_j + S^y_i S^y_j$) contributions at first order. Given the HP transformation one sees that $\mathcal{O}^{(0)}_{ij}$ comes from the longitudinal part and is simply the $(-1)^r S^2$ term in Eq.~(20). The first order term $\mathcal{O}^{(1)}_{ij}$ is made of both longitudinal and transverse parts. The on-site terms ($\aopd{i}\aop{i}$ and $\aopd{j}\aop{j}$) in Eq.~\eqref{sm::equ::Oij1} come from the $S^z_i S^z_j$ terms, which can indeed be obtained by keeping the $\mathcal{O}(S)$ terms when using Eq.~\eqref{sm::equ::HP_Sz}. Notice also that they do not depend on the distance between site $i$ and site $j$. The terms which mix $i$ and $j$ in Eq.~\eqref{sm::equ::Oij1} are the ones coming from the transverse parts. They lead, after some development, to an explicit dependence on distance $|i-j|$, as can be seen in Eq.~(21-22).

Here our aim is to derive Eq.~(23) and, for simplicity, we will consider the longitudinal and transverse parts together. The cancellation of infrared divergences between longitudinal and transverse parts explicitly shown in the main text, with the details provided above, occur precisely in the same manner at second order.

The definition of the following quantity will be handy when computing expectation values:
\be
\gamma_k(r) = \cos(k r).
\ee

We proceed now to the perturbative development. Our aim is to compute the ground state expectation value of the two-point function,
\be
C_{ij} := \braket{\bold{S}_i \cdot \bold{S}_j} = \braket{\mathcal{O}_{ij}}
\ee
using Rayleigh-Schr\"odinger perturbation theory. Developing this expectation value, we obtain
\be
C_{ij} = S^2 \left[ C_{ij}^{(0)} + \frac{1}{S} C_{ij}^{(1)} + \frac{1}{S^2} C_{ij}^{(2)} + \mathcal{O}(S^{-3}) \right]
\ee
where
\begin{align}
C_{ij}^{(0)} = & \ (-1)^{i-j}, \\
%-------------------
C_{ij}^{(1)} = & \ \braket{0|\mathcal{O}_{ij}^{(1)}|0}, \\
%-------------------
C_{ij}^{(2)} = & \ \braket{0|\mathcal{O}_{ij}^{(2)}|0} + \braket{0|\mathcal{O}_{ij}^{(1)}|\Psi_0^{(1)}} + \braket{\Psi_0^{(1)}|\mathcal{O}_{ij}^{(1)}|0}.\label{Cij2}
\end{align}
In these expressions, $\ket{0}$ is the unperturbed ground state corresponding to the Bogoliubov vacuum and $\ket{\Psi^{(1)}_0}$ is its first order correction in Rayleigh-Schr\"odinger perturbation theory, satisfying $\braket{0|\Psi^{(1)}_0} = 0$. Thanks to the absence of terms in $V_1$ with $2$ creation or $2$ annihilation Bogoliubov operators it is clear that only the first term in Eq.~\eqref{Cij2} is potentially non-vanishing. All tools have now been given and we state the results for distance $r = |i-j|$, setting the lattice spacing $a=1$ and taking the continuum limit in the last equality:\\

\textbf{Even distance $r = |i-j|$:}\\
\be
C_{ij}^{(1)} = \braket{0|\mathcal{O}_{ij}^{(1)}|0} = \delta_{r,0} + \frac{2}{N} \sum_k \left( \gamma_k(r) - 1 \right) v_k^2 = 1 - \frac{2}{\pi} J_0(r),
\label{sm::equ::Cij1_even}
\ee

\be
C_{ij}^{(2)} = \frac{1}{4} \left( C^{(1)}_{ij} - \delta_{r,0} \right)^2 = \frac{1}{4} \left( 1 - \delta_{r,0} - \frac{2}{\pi} J_0(r) \right)^2,
\ee
where the IR-convergent integral $J_0(r)$ is given by
\be
J_0(r) := \int_0^{\pi/2} \dd k \, \frac{1 - \cos(k r)}{\sin k}.
\label{equ::def_J0integral}
\ee

\textbf{Odd distance $r = |i-j|$:}

\be
C_{ij}^{(1)} = \frac{2}{N} \sum_k v_k \left( v_k - \gamma_k(r) u_k \right) = -1 + \frac{2}{\pi} J_1(r),
\label{sm::equ::Cij1_odd}
\ee

\be
C_{ij}^{(2)} = - \frac{1}{4} \left( C_{ij}^{(1)} \right)^2
\ee
where the IR-convergent integral $J_1(r)$ is given by
\be
J_1(r) := \int_0^{\pi/2} \dd k \, \left[ \frac{1}{\sin k} - \frac{\cos(k r)}{\tan k} \right].
\label{equ::def_J1integral}
\ee

Here, for the sake of simplicity, we have used Rayleigh-Schr\"odinger perturbation theory, but we could have used a different path. In particular an elegant approach is to use HP Green's functions perturbation theory. The latter is the method to favor for computing the dynamical spin structure factor at zero temperature. We will omit this approach to avoid redundancy with the finite-temperature calculation treated in Section~\ref{sec5}.

\section{Asymptotic behavior of $J_0(n)$ and $J_1(n)$\label{sec3}}

We now compute the asymptotic (large distance) behavior of the IR-convergent integrals $J_0(n)$ and $J_1(n)$.\\

\textbf{Claim:}
\be
J_0(n) = \ln 2 + \gamma + \ln n + \mathcal{O}(n^{-4}), \qquad n \ \text{even},
\label{sm::equ::J0_large_dist}
\ee
where $\gamma$ is the Euler-Mascheroni constant.\\

\textbf{Proof:}\\

We rewrite $J_0(n)$ as,
\be
\begin{aligned}
- J_0(n) = \int_0^{\pi/2} \dd k \ \frac{\cos(n k) - 1}{\sin k} = & \: \int_{0}^{\pi/2} \dd k \, \left[ \frac{1}{k} - \frac{1}{\sin k} \right] \\
+ & \: \int_0^{\pi/2} \dd k \ \cos(n k) \left[ \frac{1}{\sin k} - \frac{1}{k} \right] \\
+ & \: \int_0^{\pi/2} \dd k \  \frac{\cos(n k) - 1}{k}.
\end{aligned}
\label{int_to_compute}
\ee
The first integral in the right-hand side can be evaluated exactly,
\be
\int_{0}^{\pi/2} \dd k \, \left[ \frac{1}{k} - \frac{1}{\sin k} \right] = \ln\left( \frac{\pi}{4} \right)
\label{equ::first_integral}
\ee
while the third can be related to the Cosine Integral,
\be
\int_0^{\pi/2} \dd k \  \frac{\cos(n k) - 1}{k} = \text{Ci}(n \pi/2) - \gamma - \ln(n \pi/2)
\label{equ::third_integral}
\ee
where $\gamma$ is the Euler-Mascheroni constant. Using the asymptotic behavior of the Cosine Integral and the fact that $n$ is an even integer, we obtain,
\be
\text{Ci}(n \pi/2) = \frac{4 (-1)^{n/2+1}}{n^2 \pi^2} + \mathcal{O}(n^{-4}).
\label{equ::asympt_beh_Ci}
\ee
We focus now on the second integral in Eq.~\eqref{int_to_compute}. We define
\be
f(k) = \frac{1}{\sin k} - \frac{1}{k}
\ee
and observe that
\be
f(k) = \frac{1}{k} \left[ \frac{1}{\sinc(k)} - 1 \right]
\ee
Rewriting $\sinc(k) = 1 - \left( 1 - \sinc(k) \right)$, observing that $\sinc(k) \in [\frac{2}{\pi}, 1]$ for $k \in [0,\frac{\pi}{2}]$ and using the Taylor expansion of the $\sinc$ function we obtain
\be
\frac{1}{\sinc(k)} = \sum_{p=0}^{\infty} k^{2 p} \left[ \sum_{m=0}^{\infty} \frac{(-1)^m}{(2 m + 3)!} k^{2 m} \right]^p.
\ee
Then we have
\be
f(k) = \sum_{m=0}^{\infty} a_m k^{2 m + 1} \quad \text{for} \ k \in [0,\pi/2]
\ee
with the coefficients
\be
a_m = \frac{2 (-1)^m (2^{2 m + 1} - 1) B_{2(m+1)}}{(2 m + 2)!}
\label{equ::coeffs_a_m}
\ee
where $B_{2l}$ are the first Bernoulli numbers. We can thus evaluate the integral in the following way:
\be
\int_0^{\pi/2} \dd k \ \cos(n k) f(k) = \sum_{m=0}^{\infty} a_m \mathcal{J}_m(n)
\label{equ::integral_cos2nkfk}
\ee
where
\be
\mathcal{J}_m(n) = \int_0^{\pi/2} \dd k \ k^{2 m + 1} \cos(n k).
\ee
Integrating we get
\be
\mathcal{J}_m(n) = \frac{1}{2} \left( \frac{\pi}{2} \right)^{2 m + 2} \, \frac{1}{m + 1} \ {}_1F_2\left(\bold{a}; \bold{b}; - \frac{n^2 \pi^2}{16} \right)
\label{equ_Jnm}
\ee
with
\be
\bold{a} = m + 1, \quad \bold{b} = (\frac{1}{2}, m + 2)
\ee
and ${}_pF_q(\bold{a}; \bold{b}; z)$ is the generalized hypergeometric function. It is interesting to note that Eq.~\eqref{equ_Jnm} is valid for any value of $n$, not only even integers. The integral $\mathcal{J}_0$ for $n$ integer is given by
\be
%\begin{doublespacing}
\mathcal{J}_0(n) = 
\left\{
\begin{array}{ll}
\frac{(-1)^{n/2}-1}{n^2} & \text{for} \ n \ \text{even}, \\
- \frac{1}{n^2} + \frac{(-1)^{(n-1)/2} \pi}{2 n} & \text{for} \ n \ \text{odd}. \\
\end{array}
\right.
\label{equ::J_0}
%\end{doublespacing}
\ee
Let us now prove  the following property of $\mathcal{J}_m(n)$ \emph{for $n$ even}:
\be
|\mathcal{J}_m(n)| = \mathcal{O}\left(\frac{1}{n^2}\right) \hspace{1cm} \text{for} \ n \ \text{even}.
\ee
Assume that
\be
\exists \, \bar{m} \in \mathbb{N^*} \ \text{such that} \ \mathcal{J}_{\bar{m}-1}(n) = \mathcal{O}\left(\frac{1}{n^2} \right)
\ee
and let us prove the relation for $\mathcal{J}_{\bar{m}}$. We have
\be
\mathcal{J}_{\bar{m}} = \int_{0}^{\pi/2} \dd k \ k^{2 \bar{m} + 1} \cos(n k)
\ee
Performing integration by part twice and using the fact that $n$ is an even integer, we get
\be
\mathcal{J}_{\bar{m}}(n) = (-1)^{n/2} \frac{2 \bar{m} + 1}{n^2} \left(\frac{\pi}{2}\right)^{2 \bar{m}} - \frac{2 \bar{m} (2 \bar{m} + 1)}{n^2} \mathcal{J}_{\bar{m}-1}(n)
\label{equ::Jmbar}
\ee
which proves the statement (together with Eq.~\eqref{equ::J_0}). Bringing together Eq.~\eqref{equ::first_integral},~\eqref{equ::third_integral} and \eqref{equ::integral_cos2nkfk}, we obtain
\be
- J_0(n) = \ln\left( \frac{\pi}{4} \right) + \text{Ci}(n \pi/2) - \gamma - \ln(n \pi/2) + \sum_{m=0}^{\infty} a_m \mathcal{J}_m(n)
\ee
for $n$ even. Using Eq.~\eqref{equ::Jmbar}, we get
\be
J_0(n) = \ln 2 - \text{Ci}(n \pi/2) + \gamma + \ln n - (-1)^{n/2} \frac{1}{n^2} \, \sum_{m=0}^{\infty} a_m (2 m + 1) \left(\frac{\pi}{2}\right)^{2 m} + \mathcal{O}(n^{-4})
\ee
where the coefficients $a_m$ are given by Eq.~\eqref{equ::coeffs_a_m}. Observe now that we can actually compute explicitly the coefficient of the term proportional to $n^{-2}$. Recall that
\be
f(k) = \frac{1}{\sin(k)} - \frac{1}{k} = \sum_{m=0}^{\infty} a_m k^{2 m + 1}
\ee
leading to
\be
f'(k) = \frac{1}{k^2} - \cot(k) \csc(k) = \sum_{m=0}^{\infty} a_m (2 m + 1) k^{2 m}.
\ee
Thus we have
\be
\sum_{m=0}^{\infty} a_m (2 m + 1) \left(\frac{\pi}{2}\right)^{2 m} = f'\left(\frac{\pi}{2}\right) = \frac{4}{\pi^2}
\ee
leading to
\be
\sum_{m=0}^{\infty} a_m \mathcal{J}_m(n) = (-1)^{n/2} \frac{4}{\pi^2} \frac{1}{n^2} + \mathcal{O}(n^{-4}).
\ee
The integral $J_0(n)$ is then given by
\be
J_0(n) = \ln 2 - \text{Ci}(n \pi/2) + \gamma + \ln n - \frac{4 (-1)^{n/2}}{n^2 \pi^2}  + \mathcal{O}(n^{-4}).
\ee
The final result is obtained by replacing $\text{Ci}(n \pi/2)$ by its expression in Eq.~\eqref{equ::asympt_beh_Ci}.\\

\textbf{Claim:}
\be
J_1(n) = \ln 2 + \gamma + \ln n - \frac{1}{2 n^2} + \mathcal{O}(n^{-4}), \qquad n \ \text{odd},
\label{sm::equ::J1_large_dist}
\ee
where $\gamma$ is the Euler-Mascheroni constant.\\

\textbf{Proof:}\\

The proof above can be straightforwardly extended to this case.\\

\noindent
Inserting Eq.~\eqref{sm::equ::J0_large_dist} and~\eqref{sm::equ::J1_large_dist} in Eq.~(23) leads to Eq.~(25).

\section{Static spin structure factor at zero temperature\label{sec4}}

In this section we extract the static spin structure factor at zero temperature. The easiest way to proceed is to explicitly Fourier transform the static spin-spin correlation function. This is actually a trivial task since the static spin-spin correlation function in real space was expressed as a Fourier transform. We thus only need to rewrite the expressions in a convenient way. Proceeding order by order we begin with Eq.~\eqref{sm::equ::Cij1_even} and~\eqref{sm::equ::Cij1_odd}, remove the distance-independent terms and make the Fourier transform manifest. We thus write
\be
C_{e}^{(1),xx}(r) = \frac{1}{2} \frac{1}{N} \sum_k (u_k^2 + v_k^2) e^{i k r} \quad \text{(even distance)},
\label{equ::Sk::xx::Ce1}
\ee
\be
C_{o}^{(1),xx}(r) = - \frac{1}{N} \sum_k u_k v_k e^{i k r} \quad \text{(odd distance)}.
\label{equ::Sk::xx::Co1}
\ee
In order to implement the even-odd alternation in a single equation, we write
\be
C^{(1),xx}(r) = (-1)^r \frac{C_{e}^{(1),xx}(r) - C_{o}^{(1),xx}(r)}{2} + \frac{C_{e}^{(1),xx}(r) + C_{o}^{(1),xx}(r)}{2}
\ee
which can be reexpressed as
\be
C^{(1),xx}(r) = \frac{1}{4} \left[ (-1)^r \frac{1}{N} \sum_k (u_k + v_k)^2 e^{i k r} + \frac{1}{N} \sum_k (u_k - v_k)^2 e^{i k r} \right].
\ee
Defining $Q = \pi$ the pitch vector we have $(-1)^r = e^{i Q r}$ and the previous equation becomes
\be
C^{(1),xx}(r) = \frac{1}{4} \left[ \frac{1}{N} \sum_k (u_k + v_k)^2 e^{i (k+Q) r} + \frac{1}{N} \sum_k (u_k - v_k)^2 e^{i k r} \right].
\ee
The transverse part of the equal-time structure factor at first order can be extracted after little algebra:
\be
S^{xx}(k) = \frac{S}{2} \left|  \tan\left(\frac{k}{2}\right)\right|.
\ee
Proceeding similarly for the second order terms one finally obtains
\be
S^{xx}(k) = \frac{S}{2} \left( 1 - \frac{n}{S} \right) \left|  \tan\left(\frac{k}{2}\right)\right|
\label{sm::equ::Sxxk}
\ee
where $n$ is the IR-divergent coefficient in Eq.~\eqref{equ::def_n}.\\

The longitudinal part is treated similarly. One obtains
\be
S^{zz}(k) = - \frac{1}{4} + \frac{1}{4\pi} \int_0^{\pi} \dd q \, \frac{1 - \cos q \cos(k+q)}{\sin q |\sin(k+q)|}
\label{equ::Sk::Szzk_final}
\ee
which is a divergent integral for any non-zero value of $k$. However it is a simple task to show that the O($3$) invariant structure factor $S(k) = 2 S^{xx}(k) + S^{zz}(k)$ is convergent to second order and is given by
\be
S(k) = \frac{2 S + 1}{2} \left|\tan\left(\frac{k}{2}\right)\right| - \frac{1}{4} + \frac{1}{4\pi} \int_0^{\pi} \dd q \, \frac{1}{\sin q} \left[ \frac{1 - \cos q \cos(k+q)}{|\sin(k+q)|} - 2 |\tan(k/2)| \right]
\ee
where we used the definition of the divergent coefficient $n$ given in Eq.~\eqref{equ::def_n}.

\section{Dynamical spin structure factor at finite temperature\label{sec5}}

We now turn to the calculation of the dynamical spin structure factor at finite temperature $1/\beta$. The zero-temperature results can be obtained as the $\beta\rightarrow\infty$ limit of our final results. We begin by defining the imaginary-time ordered Green's function of Holstein-Primakoff bosons, which is nothing but the extension of Eq.~(26) to finite temperature,
\be
\mathbb{G}(k,\tau) = - \braket{T_{\tau} \begin{bmatrix}
\aop{k}(\tau) \\ \aopd{-k}(\tau)
\end{bmatrix}
\begin{bmatrix}
\aopd{k}(0) & \aop{-k}(0) \\
\end{bmatrix}}
=- \begin{pmatrix}
\braket{T_{\tau} \aop{k}(\tau)\aopd{k}(0)} & \braket{T_{\tau} \aop{k}(\tau)\aop{-k}(0)} \\
\braket{T_{\tau} \aopd{-k}(\tau)\aopd{k}(0)} & \braket{T_{\tau} \aopd{-k}(\tau)\aop{-k}(0)} \\
\end{pmatrix}
\ee
where $T_{\tau}$ is the imaginary-time ordering operator. Taking the expectation values in the unperturbed ground state leads to the free Green's function,
\be
\mathbb{G}^0(k,i\omega_n) = \frac{1}{(i\omega_n)^2 - \epsilon_k^2}
\begin{pmatrix}
A_k + i\omega_n & B_k \\
B_k & A_k - i \omega_n \\
\end{pmatrix}
\ee
where $\omega_n = 2 n \pi/\beta, \ n \in \mathbb{Z}$ are the bosonic Matsubara frequencies and where
\be
\mathbb{G}^0(k,i\omega_n) = \int_0^{\beta} \dd \tau \, e^{i\omega_n \tau} \mathbb{G}^0(k,\tau).
\ee
We perform now a standard (but somehow tedious) perturbative expansion of these Green's functions by incorporating the first order interaction term of the Hamiltonian and then use Dyson's equation to rewrite the Green's function as
\be
\mathbb{G}(k,i\omega_n) = \frac{1}{(i \omega_n)^2 - \xi_k^2} 
\begin{pmatrix}
A_k \left( 1 - \frac{\kappa_{\beta}}{S} \right) + i \omega_n & B_k \left( 1 - \frac{\kappa_{\beta}}{S} \right) \\
B_k \left( 1 - \frac{\kappa_{\beta}}{S} \right) & A_k \left( 1 - \frac{\kappa_{\beta}}{S} \right) - i \omega_n \\
\end{pmatrix}
\label{equ::HP_finiteT::G_final_freq}
\ee
where
\be
\xi_k = \left( 1 + \frac{|\kappa_{\beta}|}{S} \right) \epsilon^{(0)}_k = 2 J S  \left( 1 + \frac{|\kappa_{\beta}|}{S} \right) |\sin k| \equiv v_{\beta} |\sin k|
\ee
is the NLSWT dispersion relation and has a first order correction which depends on the inverse temperature $\beta$ through
\be
\kappa_{\beta} = \frac{1}{N} \sum_q \frac{|\sin q| \coth\left(\frac{\beta \epsilon^{(0)}_q}{2} \right) - 1}{2}.
\label{equ::kappa_beta_developped}
\ee
This is the obvious extension of Eq.~\eqref{equ::def_kappa} to finite temperature. Note that $\kappa_{\beta} \rightarrow \kappa$ as $\beta\rightarrow\infty$.\\

We proceed now to the actual calculation of the dynamical structure factor. We focus first on the longitudinal part. We define the following time-ordered spin two-point function at finite temperature in imaginary time,
\be
F^{zz}(k,i\omega_n;\beta) = - \int_0^{\beta} \dd \tau \, e^{i \omega_n \tau} \braket{T_{\tau} S^z_k(\tau) S^z_{-k}(0)}
\label{equ::DSFT::Fzz_imagtime}
\ee
where $S^z_k(\tau)$ is an imaginary-time evolved spin operator $S^z$ in Fourier space. Performing an analytic continuation gives the real-frequency, retarded two-point function,
\be
\lim_{i\omega_n \rightarrow \omega + i\eta} F^{zz}(k,i\omega_n;\beta) = F^{zz,R}(k,\omega;\beta)
\ee
from which we can obtain the dynamical spin structure factor at finite temperature $1/\beta$,
\be
S^{zz}(k,\omega;\beta) = - \frac{2}{1 - e^{-\beta \omega}} \, \text{Im}\left[ F^{zz,R}(k,\omega;\beta) \right].
\label{equ::DSFT::SzzT_from_Fzz}
\ee
In practice we will compute $\tilde{F}^{zz}(k,i\omega_n;\beta)$ which is the equivalent of Eq.~\eqref{equ::DSFT::Fzz_imagtime} but for the rotated spin operators and will shift momentum by $\pi$ at the very end of the calculation. We now provide the reader with intermediary steps. We begin by expanding the spin operators in HP bosons and keep only the relevant $2$-magnon terms. Using Wick's theorem and performing the $\tau$ integral we end up with
\be
F^{zz}(k, i\omega_n; \beta) = - \frac{1}{N} \sum_q \frac{1}{\beta} \sum_m \left[ \mathbb{G}_{12}(q-k,i\omega_n-i\omega_m) \, \mathbb{G}_{12}(q,i\omega_m) + \mathbb{G}_{22}(q-k,i\omega_n-i\omega_m) \, \mathbb{G}_{11}(q,i\omega_m) \right].
\label{equ::DSFT::Fzz2magnon_1}
\ee
Performing the Matsubara sums, we get
\be
- \tilde{F}^{zz}(k,i\omega_n) =  \frac{1}{N} \sum_q \left[ P_{k,q} - i\omega_n Q_{k,q} \right] S_1(k,q,i\omega_n) 
%------------
+  \ \frac{1}{N} \sum_q \left( R_{k,q} - i \omega_n \right) S_2(k,q,i\omega_n) 
%------------
+  \ \frac{1}{N} \sum_q S_3(k,q,i\omega_n)
\label{equ::DSFT::Fzz_final_imaginaryfrequency}
\ee
where
\begin{align}
P_{k,q} = & \ \left( 1 - \frac{\kappa_{\beta}}{S} \right)^2 ( A_q A_{k-q} + B_q B_{k-q} ), \\
%----------
Q_{k,q} = & \ \left( 1 - \frac{\kappa_{\beta}}{S} \right) A_q, \\
%----------
R_{k,q} = & \ \left( 1 - \frac{\kappa_{\beta}}{S} \right) (A_q + A_{q-k}),
\end{align}
and where the Matsubara sums $S_1$, $S_2$ and $S_3$ take the following expressions:
\be
S_1(k,q,i\omega_n) = \frac{1}{\beta} \sum_m \frac{1}{(i\omega_m)^2 - \xi_q^2} \, \frac{1}{(i\omega_m-i\omega_n)^2 - \xi_{k-q}^2},
\label{equ::DSFT::S1_def}
\ee

\be
S_2(k,q,i\omega_n) = \frac{1}{\beta} \sum_m \frac{i\omega_m}{(i\omega_m)^2 - \xi_q^2} \, \frac{1}{(i\omega_m-i\omega_n)^2 - \xi_{k-q}^2},
\label{equ::DSFT::S2_def}
\ee

\be
S_3(k,q,i\omega_n) = \frac{1}{\beta} \sum_m \frac{(i\omega_m)^2}{(i\omega_m)^2 - \xi_q^2} \, \frac{1}{(i\omega_m-i\omega_n)^2 - \xi_{k-q}^2}.
\label{equ::DSFT::S3_def}
\ee
The next step is the analytic continuation of this expression. This is a lengthy but almost trivial calculation which makes use of Sokhotsky's formula:
\be
\frac{1}{x \pm i \eta} = \text{PV} \frac{1}{x} - i\pi \delta(x).
\label{sm::equ::sokhotsky}
\ee
We end up with our final expression of the dynamical spin structure factor at finite temperature $1/\beta$:
\be
\begin{aligned}
S^{zz}(k,\omega;\beta) = & \ \frac{2\pi}{1 - e^{-\beta\omega}} \, \frac{1}{N} \sum_{q>0} \left[ T^{(1)}_{k,q} \delta(\omega - \xi_q - \xi_{k+q}) + T^{(2)}_{k,q} \delta(\omega + \xi_q + \xi_{k+q}) \right. \\
%--------
& \hspace{5cm} \left. + T^{(3)}_{k,q} \delta(\omega - \xi_q + \xi_{k+q}) + T^{(4)}_{k,q} \delta(\omega + \xi_q - \xi_{k+q}) \right]
\end{aligned}
\label{equ::DSFT::Szzfinal_finiteT_1}
\ee
where
\be
T_{k,q}^{(1)} = \frac{1}{4} \left[ \coth\left(\frac{\beta\xi_q}{2}\right) + \coth\left(\frac{\beta\xi_{k+q}}{2}\right) \right] \left[ \frac{1 - \cos q\cos(k+q) + \sin q - |\sin(k+q)|}{\sin q \, |\sin(k+q)|} - 1 \right],
\label{equ::DSFT::Tkq1_final}
\ee

\be
T_{k,q}^{(2)} = - \frac{1}{4} \left[ \coth\left(\frac{\beta\xi_q}{2}\right) + \coth\left(\frac{\beta\xi_{k+q}}{2}\right) \right] \left[ \frac{1 - \cos q \cos(k+q) - \sin q + |\sin(k+q)|}{\sin q\, |\sin(k+q)|} - 1 \right],
\label{equ::DSFT::Tkq2_final}
\ee

\be
T_{k,q}^{(3)} = - \frac{1}{4} \left[ \coth\left(\frac{\beta\xi_q}{2}\right) - \coth\left(\frac{\beta\xi_{k+q}}{2}\right) \right] \left[ \frac{1 - \cos q \cos(k+q) + \sin q + |\sin(k+q)|}{\sin q\, |\sin(k+q)|} + 1 \right],
\label{equ::DSFT::Tkq3_final}
\ee

\be
T_{k,q}^{(4)} = \frac{1}{4} \left[ \coth\left(\frac{\beta\xi_q}{2}\right) - \coth\left(\frac{\beta\xi_{k+q}}{2}\right) \right] \left[ \frac{1 - \cos q \cos(k+q) - \sin q - |\sin(k+q)|}{\sin q\, |\sin(k+q)|} + 1 \right].
\label{equ::DSFT::Tkq4_final}
\ee
Equation~\eqref{equ::DSFT::Szzfinal_finiteT_1} is plotted in Fig.~2 for spin $S = 5/2$ and inverse temperature $\beta = 10$ (in units of $J^{-1}$).\\

Having worked out the longitudinal part we focus now on the transverse dynamical spin structure factor. We define $F^{xx}(k, i\omega_n; \beta)$ in a similar fashion as in Eq.~\eqref{equ::DSFT::Fzz_imagtime}. We obtain
\be
\tilde{F}^{xx}(k,i\omega_n; \beta) = \frac{S}{2} \left( 1 - \frac{n_{\beta}}{S} \right) \sum_{j,l=1,2} \mathbb{G}_{j,l}(k,i\omega_n)
\ee
where
\be
n_{\beta} = - \frac{1}{N} \sum_k \mathbb{G}^0_{22}(k,\tau=0^{+}) = \frac{1}{N} \sum_k \frac{u_k^2 + v_k^2 e^{\beta\epsilon_k}}{e^{\beta\epsilon_k} - 1} \qquad \text{(IR-divergent)}
\ee
is the equivalent of Eq.~\eqref{equ::def_n} at finite temperature. The next steps, analytical continuation and application of the equivalent of Eq.~\eqref{equ::DSFT::SzzT_from_Fzz}, respectively, follow straightforwardly. We end up with
\be
S^{xx}(k,\omega;\beta) = \frac{S \pi  \text{sign}(\omega)}{1 - e^{-\beta \omega}} \, \left( 1 - \frac{n_{\beta}}{S} \right) \left|\tan\left(\frac{k}{2}\right)\right| \, \left( \delta(\omega-\xi_k) + \delta(\omega+\xi_k) \right).
\label{equ::DSFT::transverse_final}
\ee

\section{Zero-temperature dynamical spin structure factor\label{sec6}}

In the $\beta\rightarrow\infty$ limit the $T^{(1)}_{k,q}$ term in Eq.~\eqref{equ::DSFT::Szzfinal_finiteT_1} is the only surviving contribution and corresponds precisely to the result that can be obtained by the standard zero-temperature pertubative theory of real-time Green's functions. We obtain
\be
S^{zz}(k,\omega) = \lim_{\beta\to\infty} S^{zz}(k,\omega;\beta) = \frac{1}{2} \int_0^{\pi} \dd q \, \left[ \frac{1 - \cos q \cos(k+q)}{\sin q\, |\sin(k+q)|} - 1 \right] \, \delta(\omega - \epsilon_q - \epsilon_{k+q})
\ee
where $\epsilon_q$ is the NLSWT dispersion relation at zero temperature given in Eq.~\eqref{sm::equ::epsilonk_nlswt} and we have taken the continuum limit. For consistency check one can verify that integration of this expression over positive frequency using Eq.~(30) indeed leads to Eq.~\eqref{equ::Sk::Szzk_final}.

Similarly taking the $\beta\rightarrow\infty$ limit of Eq.~\eqref{equ::DSFT::transverse_final} leads to
\be
S^{xx}(k,\omega) = S \pi \left( 1 - \frac{n}{S} \right) \left|\tan\left(\frac{k}{2}\right)\right| \, \delta(\omega-\epsilon_k)
\ee
which is precisely Eq.~(29). It is straighforward to show that applying Eq.~(30) to $S^{xx}(k,\omega)$ one gets back Eq.~\eqref{sm::equ::Sxxk}.

\bibliography{references}
\bibliographystyle{apsrev4-1}